\documentclass[final]{IEEEtran}
\usepackage[update,prepend]{epstopdf}
\usepackage{todonotes}
\usepackage{graphics}
\usepackage{multirow}
\usepackage{tikz}
\usepackage{bbm} 
\usepackage{pdfpages}
\usepackage{multirow}
\usepackage{tabularx}
\usepackage{subfigure}
\usepackage{comment}
\usepackage{setspace}	
\usepackage{graphicx}
\usepackage{algorithm,algorithmic}
\usepackage{multicol}

\usepackage[justification=centering]{caption}
\usepackage{textcomp}
\usepackage{psfrag}
\usepackage{arydshln}
\usepackage{url}
\usepackage{soul}
\usepackage{graphicx,color}
\usepackage[nolist]{acronym}

\usepackage{mathtools,lipsum}
\usepackage{cuted}
\usepackage{amsmath}
\usepackage{graphicx}

\newcommand{\revise}[1]{\textcolor{black}{#1}}

\def\nb0{{\mathbf{0}}}
\def\nb1{{\mathbf{1}}}

\begin{document}
\graphicspath{{./Figures/}}
	\begin{acronym}

\acro{5G-NR}{5G New Radio}
\acro{3GPP}{3rd Generation Partnership Project}
\acro{ABS}{aerial base station}
\acro{AC}{address coding}
\acro{ACF}{autocorrelation function}
\acro{ACR}{autocorrelation receiver}
\acro{ADC}{analog-to-digital converter}
\acrodef{aic}[AIC]{Analog-to-Information Converter}     
\acro{AIC}[AIC]{Akaike information criterion}
\acro{aric}[ARIC]{asymmetric restricted isometry constant}
\acro{arip}[ARIP]{asymmetric restricted isometry property}

\acro{ARQ}{Automatic Repeat Request}
\acro{AUB}{asymptotic union bound}
\acrodef{awgn}[AWGN]{Additive White Gaussian Noise}     
\acro{AWGN}{additive white Gaussian noise}
\acro{VHeNET}{Vertical Heterogeneous Network}
\acro{LAP}{Low altitude platform}

\acro{APSK}[PSK]{asymmetric PSK} 

\acro{waric}[AWRICs]{asymmetric weak restricted isometry constants}
\acro{warip}[AWRIP]{asymmetric weak restricted isometry property}
\acro{BCH}{Bose, Chaudhuri, and Hocquenghem}        
\acro{BCHC}[BCHSC]{BCH based source coding}
\acro{BEP}{bit error probability}
\acro{BFC}{block fading channel}
\acro{BG}[BG]{Bernoulli-Gaussian}
\acro{BGG}{Bernoulli-Generalized Gaussian}
\acro{BPAM}{binary pulse amplitude modulation}
\acro{BPDN}{Basis Pursuit Denoising}
\acro{BPPM}{binary pulse position modulation}
\acro{BPSK}{Binary Phase Shift Keying}
\acro{BPZF}{bandpass zonal filter}
\acro{BSC}{binary symmetric channels}              
\acro{BU}[BU]{Bernoulli-uniform}
\acro{BER}{bit error rate}
\acro{BS}{base station}
\acro{BW}{BandWidth}
\acro{BLLL}{ binary log-linear learning }

\acro{CP}{Cyclic Prefix}
\acrodef{cdf}[CDF]{cumulative distribution function}   
\acro{CDF}{Cumulative Distribution Function}
\acrodef{c.d.f.}[CDF]{cumulative distribution function}
\acro{CCDF}{complementary cumulative distribution function}
\acrodef{ccdf}[CCDF]{complementary CDF}               
\acrodef{c.c.d.f.}[CCDF]{complementary cumulative distribution function}
\acro{CD}{cooperative diversity}

\acro{CDMA}{Code Division Multiple Access}
\acro{ch.f.}{characteristic function}
\acro{CIR}{channel impulse response}
\acro{cosamp}[CoSaMP]{compressive sampling matching pursuit}
\acro{CR}{cognitive radio}
\acro{cs}[CS]{compressed sensing}                   
\acrodef{cscapital}[CS]{Compressed sensing} 
\acrodef{CS}[CS]{compressed sensing}
\acro{CSI}{channel state information}
\acro{CCSDS}{consultative committee for space data systems}
\acro{CC}{convolutional coding}
\acro{Covid19}[COVID-19]{Coronavirus disease}

\acro{DAA}{detect and avoid}
\acro{DAB}{digital audio broadcasting}
\acro{DCT}{discrete cosine transform}
\acro{dft}[DFT]{discrete Fourier transform}
\acro{DR}{distortion-rate}
\acro{DS}{direct sequence}
\acro{DS-SS}{direct-sequence spread-spectrum}
\acro{DTR}{differential transmitted-reference}
\acro{DVB-H}{digital video broadcasting\,--\,handheld}
\acro{DVB-T}{digital video broadcasting\,--\,terrestrial}
\acro{DL}{DownLink}
\acro{DSSS}{Direct Sequence Spread Spectrum}
\acro{DFT-s-OFDM}{Discrete Fourier Transform-spread-Orthogonal Frequency Division Multiplexing}
\acro{DAS}{Distributed Antenna System}
\acro{DNA}{DeoxyriboNucleic Acid}

\acro{EC}{European Commission}
\acro{EED}[EED]{exact eigenvalues distribution}
\acro{EIRP}{Equivalent Isotropically Radiated Power}
\acro{ELP}{equivalent low-pass}
\acro{eMBB}{Enhanced Mobile Broadband}
\acro{EMF}{ElectroMagnetic Field}
\acro{EU}{European union}
\acro{EI}{Exposure Index}
\acro{eICIC}{enhanced Inter-Cell Interference Coordination}

\acro{FC}[FC]{fusion center}
\acro{FCC}{Federal Communications Commission}
\acro{FEC}{forward error correction}
\acro{FFT}{fast Fourier transform}
\acro{FH}{frequency-hopping}
\acro{FH-SS}{frequency-hopping spread-spectrum}
\acrodef{FS}{Frame synchronization}
\acro{FSsmall}[FS]{frame synchronization}  
\acro{FDMA}{Frequency Division Multiple Access}

\acro{GA}{Gaussian approximation}
\acro{GF}{Galois field }
\acro{GG}{Generalized-Gaussian}
\acro{GIC}[GIC]{generalized information criterion}
\acro{GLRT}{generalized likelihood ratio test}
\acro{GPS}{Global Positioning System}
\acro{GMSK}{Gaussian Minimum Shift Keying}
\acro{GSMA}{Global System for Mobile communications Association}
\acro{GS}{ground station}
\acro{GMG}{ Grid-connected MicroGeneration}

\acro{HAP}{high altitude platform}
\acro{HetNet}{Heterogeneous network}

\acro{IDR}{information distortion-rate}
\acro{IFFT}{inverse fast Fourier transform}
\acro{iht}[IHT]{iterative hard thresholding}
\acro{i.i.d.}{independent, identically distributed}
\acro{IoT}{Internet of Things}                      
\acro{IR}{impulse radio}
\acro{lric}[LRIC]{lower restricted isometry constant}
\acro{lrict}[LRICt]{lower restricted isometry constant threshold}
\acro{ISI}{intersymbol interference}
\acro{ITU}{International Telecommunication Union}
\acro{ICNIRP}{International Commission on Non-Ionizing Radiation Protection}
\acro{IEEE}{Institute of Electrical and Electronics Engineers}
\acro{ICES}{IEEE international committee on electromagnetic safety}
\acro{IEC}{International Electrotechnical Commission}
\acro{IARC}{International Agency on Research on Cancer}
\acro{IS-95}{Interim Standard 95}

\acro{KPI}{Key Performance Indicator}

\acro{LEO}{low Earth orbit}
\acro{LF}{likelihood function}
\acro{LLF}{log-likelihood function}
\acro{LLR}{log-likelihood ratio}
\acro{LLRT}{log-likelihood ratio test}
\acro{LoS}{line-of-sight}
\acro{LRT}{likelihood ratio test}
\acro{wlric}[LWRIC]{lower weak restricted isometry constant}
\acro{wlrict}[LWRICt]{LWRIC threshold}
\acro{LPWAN}{Low Power Wide Area Network}
\acro{LoRaWAN}{Low power long Range Wide Area Network}
\acro{NLoS}{Non-Line-of-Sight}
\acro{LiFi}[Li-Fi]{light-fidelity}
 \acro{LED}{light emitting diode}
 \acro{LABS}{LoS transmission with each ABS}
 \acro{NLABS}{NLoS transmission with each ABS}

\acro{MB}{multiband}
\acro{MC}{macro cell}
\acro{MDS}{mixed distributed source}
\acro{MF}{matched filter}
\acro{m.g.f.}{moment generating function}
\acro{MI}{mutual information}
\acro{MIMO}{Multiple-Input Multiple-Output}
\acro{MISO}{multiple-input single-output}
\acrodef{maxs}[MJSO]{maximum joint support cardinality}                       
\acro{ML}[ML]{maximum likelihood}
\acro{MMSE}{minimum mean-square error}
\acro{MMV}{multiple measurement vectors}
\acrodef{MOS}{model order selection}
\acro{M-PSK}[${M}$-PSK]{$M$-ary phase shift keying}                       
\acro{M-APSK}[${M}$-PSK]{$M$-ary asymmetric PSK} 
\acro{MP}{ multi-period}
\acro{MINLP}{mixed integer non-linear programming}

\acro{M-QAM}[$M$-QAM]{$M$-ary quadrature amplitude modulation}
\acro{MRC}{maximal ratio combiner}                  
\acro{maxs}[MSO]{maximum sparsity order}                                      
\acro{M2M}{Machine-to-Machine}                                                
\acro{MUI}{multi-user interference}
\acro{mMTC}{massive Machine Type Communications}      
\acro{mm-Wave}{millimeter-wave}
\acro{MP}{mobile phone}
\acro{MPE}{maximum permissible exposure}
\acro{MAC}{media access control}
\acro{NB}{narrowband}
\acro{NBI}{narrowband interference}
\acro{NLA}{nonlinear sparse approximation}
\acro{NLOS}{Non-Line of Sight}
\acro{NTIA}{National Telecommunications and Information Administration}
\acro{NTP}{National Toxicology Program}
\acro{NHS}{National Health Service}

\acro{LOS}{Line of Sight}

\acro{OC}{optimum combining}                             
\acro{OC}{optimum combining}
\acro{ODE}{operational distortion-energy}
\acro{ODR}{operational distortion-rate}
\acro{OFDM}{Orthogonal Frequency-Division Multiplexing}
\acro{omp}[OMP]{orthogonal matching pursuit}
\acro{OSMP}[OSMP]{orthogonal subspace matching pursuit}
\acro{OQAM}{offset quadrature amplitude modulation}
\acro{OQPSK}{offset QPSK}
\acro{OFDMA}{Orthogonal Frequency-division Multiple Access}
\acro{OPEX}{Operating Expenditures}
\acro{OQPSK/PM}{OQPSK with phase modulation}

\acro{PAM}{pulse amplitude modulation}
\acro{PAR}{peak-to-average ratio}
\acrodef{pdf}[PDF]{probability density function}                      
\acro{PDF}{probability density function}
\acrodef{p.d.f.}[PDF]{probability distribution function}
\acro{PDP}{power dispersion profile}
\acro{PMF}{probability mass function}                             
\acrodef{p.m.f.}[PMF]{probability mass function}
\acro{PN}{pseudo-noise}
\acro{PPM}{pulse position modulation}
\acro{PRake}{Partial Rake}
\acro{PSD}{power spectral density}
\acro{PSEP}{pairwise synchronization error probability}
\acro{PSK}{phase shift keying}
\acro{PD}{power density}
\acro{8-PSK}[$8$-PSK]{$8$-phase shift keying}
\acro{PPP}{Poisson point process}
\acro{PCP}{Poisson cluster process}
 
\acro{FSK}{Frequency Shift Keying}

\acro{QAM}{Quadrature Amplitude Modulation}
\acro{QPSK}{Quadrature Phase Shift Keying}
\acro{OQPSK/PM}{OQPSK with phase modulator }

\acro{RD}[RD]{raw data}
\acro{RDL}{"random data limit"}
\acro{ric}[RIC]{restricted isometry constant}
\acro{rict}[RICt]{restricted isometry constant threshold}
\acro{rip}[RIP]{restricted isometry property}
\acro{ROC}{receiver operating characteristic}
\acro{rq}[RQ]{Raleigh quotient}
\acro{RS}[RS]{Reed-Solomon}
\acro{RSC}[RSSC]{RS based source coding}
\acro{r.v.}{random variable}                               
\acro{R.V.}{random vector}
\acro{RMS}{root mean square}
\acro{RFR}{radiofrequency radiation}
\acro{RIS}{Reconfigurable Intelligent Surface}
\acro{RNA}{RiboNucleic Acid}
\acro{RRM}{Radio Resource Management}
\acro{RUE}{reference user equipments}
\acro{RAT}{radio access technology}
\acro{RB}{resource block}
\acro{SINR}{Signal-to-interference-plus-noise ratio}
\acro{SG}{stochastic geometry}

\acro{SA}[SA-Music]{subspace-augmented MUSIC with OSMP}
\acro{SC}{small cell}
\acro{SCBSES}[SCBSES]{Source Compression Based Syndrome Encoding Scheme}
\acro{SCM}{sample covariance matrix}
\acro{SEP}{symbol error probability}
\acro{SIMO}{single-input multiple-output}
\acro{SINR}{signal-to-interference plus noise ratio}
\acro{SIR}{signal-to-interference ratio}
\acro{SISO}{Single-Input Single-Output}
\acro{SMV}{single measurement vector}
\acro{SNR}[\textrm{SNR}]{signal-to-noise ratio} 
\acro{sp}[SP]{subspace pursuit}
\acro{SS}{spread spectrum}
\acro{SW}{sync word}
\acro{SAR}{specific absorption rate}
\acro{SSB}{synchronization signal block}
\acro{SR}{shrink and realign}
\acro{LOS}{line of sight}

\acro{tUAV}{tethered Unmanned Aerial Vehicle}
\acro{TBS}{terrestrial base station}

\acro{uUAV}{untethered Unmanned Aerial Vehicle}
\acro{PDF}{probability density functions}

\acro{PL}{path-loss}

\acro{TH}{time-hopping}
\acro{ToA}{time-of-arrival}
\acro{TR}{transmitted-reference}
\acro{TW}{Tracy-Widom}
\acro{TWDT}{TW Distribution Tail}
\acro{TCM}{trellis coded modulation}
\acro{TDD}{Time-Division Duplexing}
\acro{TDMA}{Time Division Multiple Access}
\acro{Tx}{average transmit}

\acro{UAV}{unmanned aerial vehicle}
\acro{uric}[URIC]{upper restricted isometry constant}
\acro{urict}[URICt]{upper restricted isometry constant threshold}
\acro{UWB}{ultrawide band}
\acro{UWBcap}[UWB]{Ultrawide band}   
\acro{URLLC}{Ultra Reliable Low Latency Communications}
         
\acro{wuric}[UWRIC]{upper weak restricted isometry constant}
\acro{wurict}[UWRICt]{UWRIC threshold}                
\acro{UE}{User Equipment}
\acro{UL}{UpLink}

\acro{WiM}[WiM]{weigh-in-motion}
\acro{WLAN}{wireless local area network}
\acro{wm}[WM]{Wishart matrix}                               
\acroplural{wm}[WM]{Wishart matrices}
\acro{WMAN}{wireless metropolitan area network}
\acro{WPAN}{wireless personal area network}
\acro{wric}[WRIC]{weak restricted isometry constant}
\acro{wrict}[WRICt]{weak restricted isometry constant thresholds}
\acro{wrip}[WRIP]{weak restricted isometry property}
\acro{WSN}{wireless sensor network}                        
\acro{WSS}{Wide-Sense Stationary}
\acro{WHO}{World Health Organization}
\acro{Wi-Fi}{Wireless Fidelity}

\acro{sss}[SpaSoSEnc]{sparse source syndrome encoding}

\acro{VLC}{Visible Light Communication}
\acro{VPN}{Virtual Private Network} 
\acro{RF}{Radio Frequency}
\acro{FSO}{Free Space Optics}
\acro{IoST}{Internet of Space Things}

\acro{GSM}{Global System for Mobile Communications}
\acro{2G}{Second-generation cellular network}
\acro{3G}{Third-generation cellular network}
\acro{4G}{Fourth-generation cellular network}
\acro{5G}{Fifth-generation cellular network}	
\acro{gNB}{next-generation Node-B Base Station}
\acro{NR}{New Radio}
\acro{UMTS}{Universal Mobile Telecommunications Service}
\acro{LTE}{Long Term Evolution}

\acro{QoS}{Quality of Service}
\end{acronym}
	
\newcommand{\SAR} {\mathrm{SAR}}
\newcommand{\WBSAR} {\mathrm{SAR}_{\mathsf{WB}}}
\newcommand{\gSAR} {\mathrm{SAR}_{10\si{\gram}}}
\newcommand{\Sab} {S_{\mathsf{ab}}}
\newcommand{\Eavg} {E_{\mathsf{avg}}}
\newcommand{\ft}{f_{\textsf{th}}}
\newcommand{\alphatf}{\alpha_{24}}

\title{System-Level Metrics for Non-Terrestrial Networks Under Stochastic Geometry Framework}

\author{
Qi Huang, {\em Student Member, IEEE}, Baha Eddine Youcef Belmekki, Ahmed M. Eltawil, {\em Senior Member}, IEEE, and Mohamed-Slim Alouini, {\em Fellow, IEEE}
\thanks{The authors are with King Abdullah University of Science and Technology (KAUST), CEMSE division, Thuwal 23955-6900, Saudi Arabia (e-mail: \{qi.huang,  bahaeddine.belmekki, ahmed.eltawil,slim.alouini\}@kaust.edu.sa.}
\vspace{-4mm}
}
\maketitle
\thispagestyle{empty}

\begin{abstract}
Non-terrestrial networks (NTNs) are considered one of the key enablers in sixth-generation (6G) wireless networks; and with their rapid growth, system-level metrics analysis adds crucial understanding into NTN system performance. 
Applying \ac{SG} as a system-level analysis tool in the context of NTN offers novel insights into the network tradeoffs. 
In this paper, we study and highlight NTN common system-level metrics from three perspectives: NTN platform types, typical communication issues, and application scenarios. 
In addition to summarizing existing research, we study the best-suited SG models for different platforms and system-level metrics which have not been well studied in the literature. In addition, we showcase NTN-dominated prospective application scenarios. Finally, we carry out a performance analysis of system-level metrics for these applications based on SG models.
\end{abstract}

\section{Introduction} 

During the last few years, the world has witnessed exponential growth in network services, especially with new use cases of sixth-generation (6G) wireless networks such as virtual reality and holographic communications. Non-terrestrial networks (NTNs) have shown excellent promise as an enabling technology for new wireless communication generations. Compared with the densely deployed terrestrial networks, NTNs are still under development, with many challenges that must be addressed and tackled such as the interconnectivity between different types of airborne platforms and terrestrial networks, and among themselves \cite{belmekki2022unleashing}.

\par
There are mainly three types of NTN platforms: \ac{LAP}, \ac{HAP}, and satellites. These platforms provide coverage for areas that the terrestrial network cannot reach. They can also enhance global connectivity with low latency and long-distance reach. Moreover, NTNs benefit from a suitable communication environment where the signal is propagated mostly through free space since it does not experience severe shadowed fading or multi-path fading. NTN platforms will be part of the 6G architecture networks. Therefore, system-level analysis, such as global connectivity, becomes more significant than the performance analysis of a single link. Unlike metrics that highlight the flying attitude and deployment of the platform, system-level metrics focus on the performance of the entire NTN system. The system-level metrics mentioned in this paper include coverage probability, availability, channel capacity, propagation, latency, and energy efficiency. 
\par

Network topology defined by the number, height, and distribution of platforms is one of the most important factors influencing system-level metrics. As a powerful mathematical tool, stochastic geometry (SG) assesses system-level metrics in large-scale network topology. Moreover, SG is also one of a few tools that can provide analytical results for system-level metrics while incorporating co-channel interference into the analysis. Although the SG-based analysis methods have powerful analytical tractability, it ignores the correlation among the platforms. For instance, the binomial point process (BPP), one of the most commonly-used SG models for satellites, assumed that satellites are distributed on the spherical surface independently and uniformly \cite{MU_2}, which differs from reality where satellites are distributed on fixed orbits. Fortunately, the SG-based system-level metrics analysis (such as coverage probability) has been proven to approximate the lower bound of the deterministic constellation, such as the Walker constellation, in the case of a large-scale satellite constellation \cite{OK_1}.
The contributions of the paper are as follows: 
\begin{itemize}
    \item While prior works focus on introducing and developing performance metrics of NTNs, this work evaluates how these metrics influence the entire NTN system performance, rather than an individual link. 
    \item To the best of the authors' knowledge, this is the first paper to specifically elucidate the system-level metrics in the NTNs, and analyze these metrics based on special specific application scenarios, including remote and rural coverage, post-disaster reconstructed networks, and military operations. Finally, the system-level metrics of these scenarios are simulated according to SG models.
    \item This paper points out new research directions on the system-level metrics of NTN platforms based on SG models, which will have a great improvement on the overall performance evaluation of the NTN network combining LAPs, HAPs, and satellites.
\end{itemize}

\par


\label{section1}
\begin{figure*}
    \centering
    \includegraphics[width=0.8\textwidth]{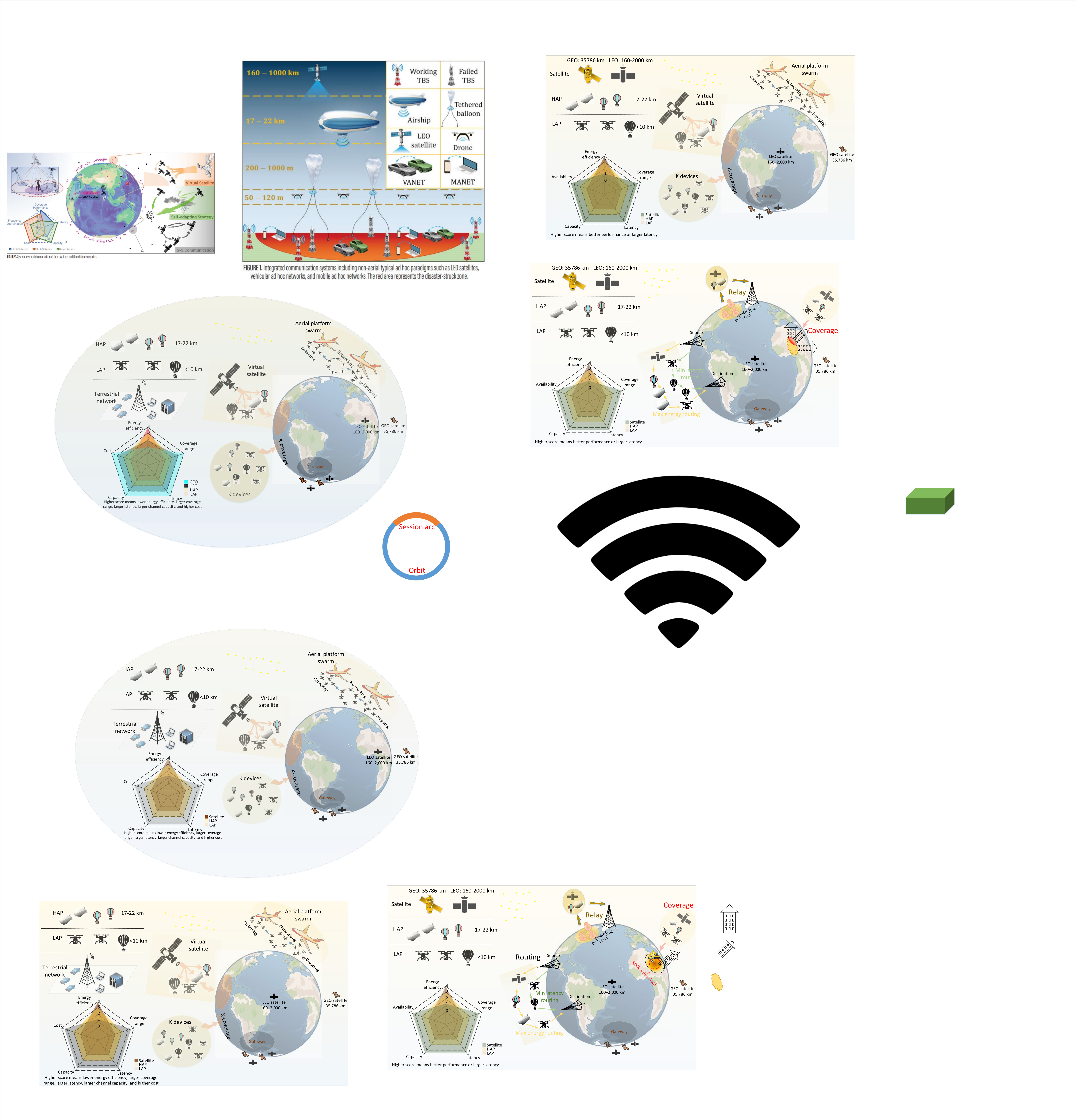}
    \caption{System framework diagram.}
    \label{syslev}
\end{figure*}

\section{NTN Platforms}\label{section2}
NTN platforms have different characteristics and features, hence, different SG models are used for each type of NTN. The radar diagram in the left bottom part of Fig.~\ref{syslev} provides the quantitative results of metrics comparison.


\subsection{Low-Altitude Platform} 
LAP refers to platforms that typically fly at 50~m -- 4~km, and applies to platforms flying under 10~km. The altitudes of most flying platforms, such as unmanned aerial vehicles (UAVs), networked tethered flying platforms (NTFPs), aerostats, and balloons, vary from tens of meters to hundreds of meters. LAPs are widely used to form a temporary network to enhance coverage owing to their flexible deployment. 
They are generally deployed in small regions at low altitudes, forming a rapidly changing topology due to their frequent movement, making the density of LAPs more important compared with their numbers. Therefore, a two-dimensional Poisson point process (PPP) is suitable for modeling a LAP network \cite{lou2021green}. 
In typical NTN application scenarios, including remote areas, post-disaster areas, and military communications (See Section~\ref{section4}), LAPs are often carried on vehicles (such as trucks) and need frequent charging, making it difficult to provide stable coverage and availability for LAPs. Moreover, the coverage area of a single LAP is significantly smaller than that of a HAP or a satellite due to its low altitude. In addition, a small channel capacity suffices to support serving targets within LAPs coverage. In comparison, LAPs possess high-energy efficiency and low propagation latency due to their lower altitude.

\subsection{High-Altitude Platform}\label{HAP}
 HAPs are flying platforms between 17 and 22~km in the stratosphere. It has not been implemented until recent years with technological progress due to the harsh environmental conditions such as low temperature, high wind speed, and long exposure to radiation. HAPs are distributed on a thin hollow spherical shell with 5~km thickness above the Earth. Therefore, HAPs are modeled approximately by BPP over a sphere. BPP is more accurate than PPP for closed surfaces such as spheres \cite{OK_1}. We note that BPP can also be used to model LAPs distributed in a finite area.

The performances of HAPs are ranked between LAPs and satellites regarding coverage area, propagation latency, channel capacity, and energy efficiency due to their altitudes which are lower than satellites and higher than LAPs. HAPs fly at high altitudes so they do not undergo frequency shadow fading caused by objects such as buildings and trees. Thus, the links between HAPs and service targets have a higher \ac{LoS} probability than LAPs. Consequently, HAPs have a stable signal-to-noise ratio (SNR) and negligible fluctuations of system-level metric values. They also benefit from higher availability (both transmitter and receiver are above the horizon) due to their greater LoS probability if deployed strategically. Higher altitude leads to better availability. Furthermore, HAPs can be powered by solar energy, hence, they can stay aloft for several years in the stratosphere.

\subsection{Low Earth Orbit Satellite}\label{LEO}

Low Earth orbit (LEO) satellites orbit around the Earth with an altitude between 160--2000~km above the Earth's surface. Examples of LEO satellite networks include Starlink and OneWeb. LEO satellites are not stationary relative to the Earth while geosynchronous equatorial orbit (GEO) satellites are geosynchronous at 35786~km. Since most GEO satellite constellations are only composed of a few satellites, SG models are not accurate and suitable for GEO satellites. For small-scale LEO satellite constellations composed of dozens of satellites (e.g., Iridium), the orbit geometry model is the closest one to approximate the actual distribution of satellites \cite{lee2022coverage}. Unlike BPP and PPP, the orbit geometry model is a semi-stochastic model, that is, the positions of orbits are deterministic, while the distribution of satellites in orbit is stochastic. As a stochastic model, details about orbits are ignored in BPP. However, spherical BPP, as is described in subsection~\ref{HAP}, is well-suited to approximate massive dense constellations. To account for the circular shape and multiple altitudes of satellite orbits, the Cox point process model was presented, where orbits are created based on PPP on a cuboid and satellites are distributed conditionally on these orbits \cite{choi2023cox}.

\par

For well-designed constellations, the unavailability of satellites on the horizon hardly occurs owing to the non-temporary nature of satellite networks. The coverage area and capacity of a single LEO or GEO satellite are much larger than those of HAPs \cite{wang2022ultra}. A single LEO satellite system provides seamless coverage for the whole Earth, while global coverage can be achieved with only three GEO satellites. The channel capacity of LEO satellite systems suffices to support communications in remote areas. The tens to hundreds of milliseconds propagation latency of LEO satellites can support most real-time communication scenarios. Satellites, including LEO and GEO, have a low energy efficiency due to the satellite-ground distance. Even with a large transmission power, the received SNR may still not exceed the decoding threshold. Due to severe attenuation in the ground clutter layer, satellite communication often requires HAPs or ground gateways for relaying.

\section{Typical Communication Analyses}\label{section3}
Depending on the number of hops, there are three communication analyses: coverage, relay, and routing. The coverage analysis using coverage probability as a metric has been the most studied system-level metric in SG literature. However, the system-level metrics in relay communications (namely availability and channel capacity) and routing (namely latency and energy efficiency) still need further investigations.


\subsection{Coverage-Based Analysis: Single-Hop Communication}
Coverage probability is a key metric in coverage-based analysis, which is the probability of received \ac{SINR} being larger than a predefined threshold. Factors that affect coverage probability include \cite{wang2022ultra}: 1) the interference power; 2) the distance to the serving platform; 3) and small-scale fading. An important feature of SG-based analysis is that the aforementioned three factors are assumed to be random variables subject to certain distributions.

\par

Interference power is the sum of the received powers of all the transmitting nodes in the network excluding the serving platform power. The authors in \cite{OK_1} show that, by taking the Laplace transform of the interference power, the specific distribution of the interference will not affect the analytical expression of the coverage probability. Second, in \cite{MU_2}, the distance to the serving platform in the SG model (contact distance distribution), is the distribution of the distance between the user and the serving platform with the strongest average received power. Finally, small-scale fading in the SG-based channel model also contributes to the randomness of received power. The air-to-ground link passing through the ground clutter layer induces a severe multi-path effect, leading to small-scale fading. Shadowed-Ricean fading is the most accurate small-scale fading model satellite to ground user link or HAP to ground \cite{jung2022performance} while that from \revise{LAP} to ground is best captured by Nakagami-$m$ fading \cite{qin2020performance}.


\subsection{Relay-Based: Dual-Hop Communication}

Relay communication can be viewed as the combination of an uplink and a downlink coverage, or a basic unit of routing. Compared with coverage and routing, availability and channel capacity are more suitable for relay-based analysis. The results of these two system-level metrics obtained in relay-based analysis provide better insight than a coverage-based one and can be easily extended to results in multi-hop routing.

\par

We start with HAP availability between two ground users. Due to blockage caused by the Earth, platforms below the horizon are out of the ground user LoS range and cannot provide a stable service. Therefore, the HAP has to be at the overlapping region composed of two spherical caps above the horizons. The area of this overlapping region is significantly smaller than a single-user LoS spherical cap when considering single-hop communications. Therefore, it is more relevant to discuss HAP availability in relay-based analysis. When HAPs are modeled as BPP, SG analysis can provide concise analytical results for HAP availability using the null probability. The SG analysis can also provide the probability of HAPs availability in the overlapping region.


\begin{table*}[ht!]
\caption{System framework diagram}
\resizebox{\textwidth}{!}{%
\begin{tabular}{|cc|c|l|l|l|c|}
\hline
\multicolumn{2}{|c|}{Platform} & Altitude & \multicolumn{1}{c|}{SG Model} & \multicolumn{1}{c|}{Advantages} & \multicolumn{1}{c|}{Challenges} & Fading model \\ \hline
\multicolumn{1}{|c|}{\multirow{4}{*}{LAP*}} & UAV & 150 m--200~m & \multirow{4}{*}{\begin{tabular}[c]{@{}l@{}}$\bullet$ PPP\\ $\bullet$ BPP for a finite area\end{tabular}} & \multirow{4}{*}{\begin{tabular}[c]{@{}l@{}}$\bullet$ High energy efficiency\\ $\bullet$ Low propagation latency\end{tabular}} & \multirow{4}{*}{\begin{tabular}[c]{@{}l@{}}$\bullet$ Limited coverage\\ area, availability, and \\ channel capacity\end{tabular}} & \multirow{4}{*}{\begin{tabular}[c]{@{}c@{}}Shadowed-Racian\\ fading\end{tabular}} \\ \cline{2-3}
\multicolumn{1}{|c|}{} & Balloon & 150 m--700~m &  &  &  &  \\ \cline{2-3}
\multicolumn{1}{|c|}{} & Blimps & 100 m--5~km &  &  &  &  \\ \cline{2-3}
\multicolumn{1}{|c|}{} & Helikites & 100 m-1.5~km &  &  &  &  \\ \hline
\multicolumn{1}{|c|}{\multirow{3}{*}{HAP}} & Aerostat & \multirow{3}{*}{\begin{tabular}[c]{@{}l@{}}~\\17 km--22~km\end{tabular}} & \multirow{3}{*}{$\bullet$ BPP} & \multirow{3}{*}{\begin{tabular}[c]{@{}l@{}}$\bullet$ More LoS probability \\ than LAP; \\ $\bullet$ Longer lifetime; Stable \\ system metrics values\end{tabular}} & \multirow{3}{*}{\begin{tabular}[c]{@{}l@{}}$\bullet$ High deployment \\ expense and difficulty; \\ $\bullet$ Low mobility\end{tabular}} & \multirow{3}{*}{\begin{tabular}[c]{@{}c@{}}Shadowed-Racian\\ fading\end{tabular}} \\ \cline{2-2}
\multicolumn{1}{|c|}{} & Airship &  &  &  &  &  \\ \cline{2-2}
\multicolumn{1}{|c|}{} & \begin{tabular}[c]{@{}c@{}}Balloon\\~\end{tabular} &  &  &  &  &  \\ \hline
\multicolumn{1}{|c|}{\multirow{2}{*}{Satellite}} & LEO & 160 km--2000~km & \multirow{3}{*}{\begin{tabular}[c]{@{}l@{}}$\bullet$ BPP \\ $\bullet$ Semi-stochastic model \\$\bullet$ Cox point process\end{tabular}} & \multirow{2}{*}{\begin{tabular}[c]{@{}l@{}}$\bullet$ Large coverage area \\ and availability\end{tabular}} & \multirow{2}{*}{\begin{tabular}[c]{@{}l@{}}$\bullet$ Long propagation latency; \\ $\bullet$ Low energy efficiency; \\ $\bullet$ Severe attenuation\end{tabular}} & \multirow{2}{*}{\begin{tabular}[c]{@{}c@{}}Nakagami-m\\ fading\end{tabular}} \\ \cline{2-3}
\multicolumn{1}{|c|}{} & \begin{tabular}[c]{@{}c@{}}GEO\\~\end{tabular} & 35786~km &  &  &  &  \\ \hline
\end{tabular}%
}
    * \footnotesize All LAP systems can be tethered to the ground \cite{belmekki2022unleashing}. 
\end{table*}

In terms of satellite system availability, we focus on LEO satellites that are not geosynchronous. LEO satellite availability is called pass duration. Pass duration is the continuous time interval when the LEO satellite is serving a given user. Denoting the arc drawn by the satellite during this interval as the session arc, the duration can be given by a fraction of the orbital period, which is equal to the ratio of the session arc length to the orbit length \cite{al2021session}. The SG-based framework provides tools to derive analytical expressions of the session arc length and pass duration. Finally, similar to the concept of LEO satellite pass duration, LAP availability is defined as the ratio of serving time to the whole period, since LAPs need to be recharged frequently. The whole period is the summation of serving time, back-haul duration, and charging duration \cite{qin2020performance}. 
\par

Channel capacity is defined as the ergodic capacity given by the Shannon-Hartley theorem over a fading channel \cite{OK_1}. Generally, a large packet is decomposed into many small packets. The channel capacity in a multi-hop link is determined by the hop with the lowest capacity. In addition, since the channel capacity in uplink and downlink might be different, it is more relevant to investigate channel capacity in relay-based analysis. Under limited transmitting power, the SG-based method maximizes channel capacity by selecting platforms at optimal locations. To make the analysis of channel capacity tractable, the SG-based method often selects a suboptimal relay strategy \cite{belbase2018coverage}. Such suboptimal relay strategy selects the platform with the maximum receiver capacity as the relay among a subset of platforms that has a communication capacity exceeding a given threshold.

\subsection{Routing-Based Analysis: Multi-Hop Communication}
Path planning and relay selections are the main focus when conducting a routing-based analysis \cite{wang2022stochastic}. Unlike relay selection in channel capacity analysis, path planning and relay selection in routing is based on all links of the entire network. When selecting a relay platform, the best relay locations for the next few hops have to be considered in advance. 
\par

Because each hop has an equal effect on the total latency, propagation latency is a typical system-level metric considered in routing. Propagation latency is defined as the time spent by the signal to propagate through the transmission medium. Propagation latency in densely distributed LAP networks has been comprehensively studied using SG framework \cite{haenggi2005routing}. As for the propagation latency as a system-level metric in HAP and satellite networks, SG is the only mathematical tool providing analytical results for spherical routing. Through contact distance distribution, the authors in \cite{wang2022stochastic} provide accurate lower and upper bounds for propagation latency in multi-hop routing.
\par

Energy efficiency is the average number of bits that can be transmitted per unit of energy consumed, which can be measured as the ratio of channel capacity to total transmission power. Generally, there are two main strategies of path and relay selection: minimum propagation latency routing and maximum-efficiency routing. The minimum propagation latency routing uses the least number of hops. Smaller number of hops leads to lower latency, but with the same transmission distance, it has a lower energy efficiency. In comparison, larger number of hops leads to larger energy efficiency since the transmission of each bit requires less energy, but the latency is larger when the transmission distance is the same.


\section{Application Scenarios}\label{section4}
In this section, we first describe three relevant application scenarios of NTNs under the SG framework.
Then, we select the most relevant system-level metrics for each application according to its specific requirement. Finally, we put forward the prospect of future research directions under the SG framework.


\subsection{Remote and Rural Coverage}
The cost of laying terrestrial infrastructures, such as optical fibers, is high. Remote areas, such as mountain areas, have sparse populations and usually low income compared to urban populations. Consequently, these areas are lacking economic incentives for mobile network operators (MNOs) to deploy terrestrial infrastructure.

\par

The most crucial communication aspects in remote areas are coverage and access equity. Therefore, investigating these two aspects is of utmost importance when designing an NTN system in these areas. Covering a large remote area with a limited number of devices at a manageable cost is challenging. Hence, studying the coverage probability under different platform distributions is necessary to design a robust NTN system. Furthermore, access equity in remote areas ensures fairness among all the users in the network. Access equity can be measured by availability, i.e., ensuring that at least one platform is available to access the network via an NTN platform.

\subsection{Post-disaster Reconstructed Networks}
When a disaster destroys terrestrial communication infrastructures, NTNs play a pivotal role in providing coverage for the disaster area. Considering that, satellite networks can achieve seamless and ubiquitous global coverage; the post-disaster network will utilize satellites as the primary NTN platform for communication. LAPs and HAPs can form a temporary network in a short period of time to increase the capacity of the post-disaster network.

\par

When the original communication networks are destroyed by a disaster, the affected areas will face a high demand for communication services in a short period. Therefore, the channel capacity of NTN is a key system-level metric to assess the quality of communications in post-disaster scenarios. Furthermore, satellites rely on solar power for recharging, which limits the amount of energy they can store. On the other hand, HAPs and LAPs require frequent recharging, which is challenging in post-disaster areas. Therefore, improving energy efficiency is critical in post-disaster scenarios.

\subsection{Military Operations}
During military operations, communication and surveillance are of the utmost importance and play a decisive role in warfare. Communication consists of the transmission of commands, orders, and instructions to their soldiers. It also consists of the transmission of information gathered through surveillance. Surveillance, on the other hand, consists of persistent and prolonged monitoring of a target or a region. Communication through terrestrial networks especially in an enemy-occupied area is difficult and dangerous during military actions. This is because the network can be easily compromised, hijacked, or eavesdropped on. In addition, surveillance requires visual observation and can be conducted using soldiers on the ground or a UAV. NTNs are a preferable alternative for terrestrial networks for this type of operation. NTNs have a larger coverage than terrestrial networks, perform surveillance with a large visual range, are easy and fast to deploy during a military operation, and have a higher LoS probability.
\par
When monitoring fast-moving military equipment, a low propagation latency is necessary so that the target will not leave the detection range, making it critical to form a high-quality network.  In addition, high bit rate images and videos help military commanders complete military deployments, take strategic decisions, and locate enemy forces. These high-resolution images or video streams rely heavily on high channel capacity. When the military has to transmit messages, orders, or reports (both in the field and at sea) or is faced with a demanding surveillance mission, the $k$-coverage probability is a more pertinent system-level metric than the coverage probability for these types of operations. The $k$-coverage is achieved when an object or region is covered by at least $k$ platforms at the same time \cite{keeler2013sinr}. NTN platforms have advantageous anti-reconnaissance ability and mobility compared with the terrestrial network platforms to form the high-quality network and satisfy the $k$-coverage probability to reliably support military operations.

\section{Simulation Results} \label{section5}
In this section, we provide comprehensive simulation results that include several system-level metrics, NTN platform types, communication analysis, and application scenarios. The models used to characterize NTN platforms and users are based on the existing SG models. To enhance the richness of the simulation results, we choose communication analysis with system-level metrics that have not been well-investigated in the literature. In the simulations, we consider decoding and forward relaying scheme, and the power consumption at each node is not considered.
\par

\subsection{Relay Communications Availability in Remote Areas} \label{simulation1}

\begin{figure}[h]
	\centering
	\includegraphics[width=0.8\linewidth]{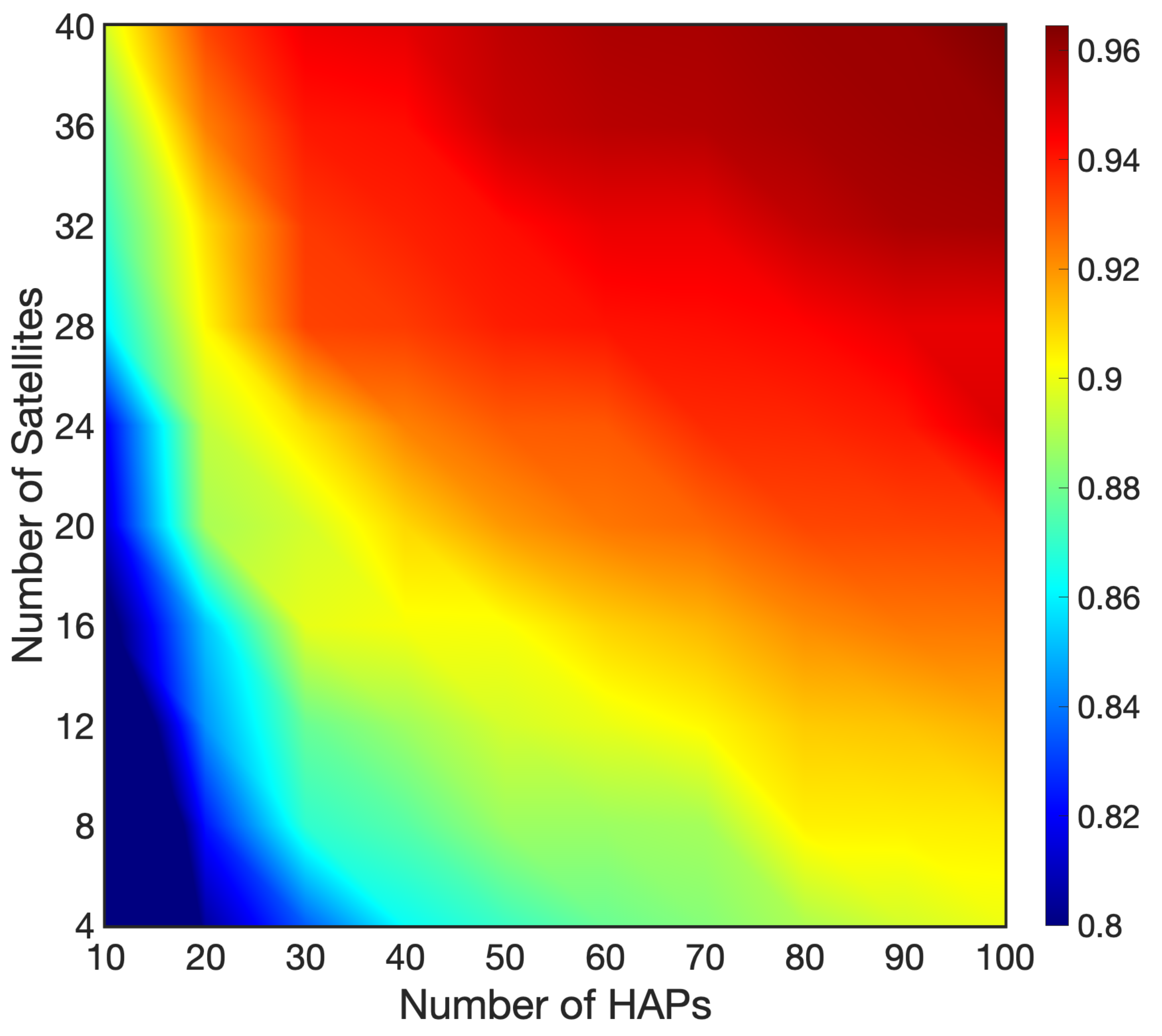}
	\caption{Availability analysis of relay communications in remote areas.}
	\label{fig:Remote}
\end{figure}


\begin{figure}
    \centering
    \subfigure[Capacity]{
    \includegraphics[width=0.43\linewidth]{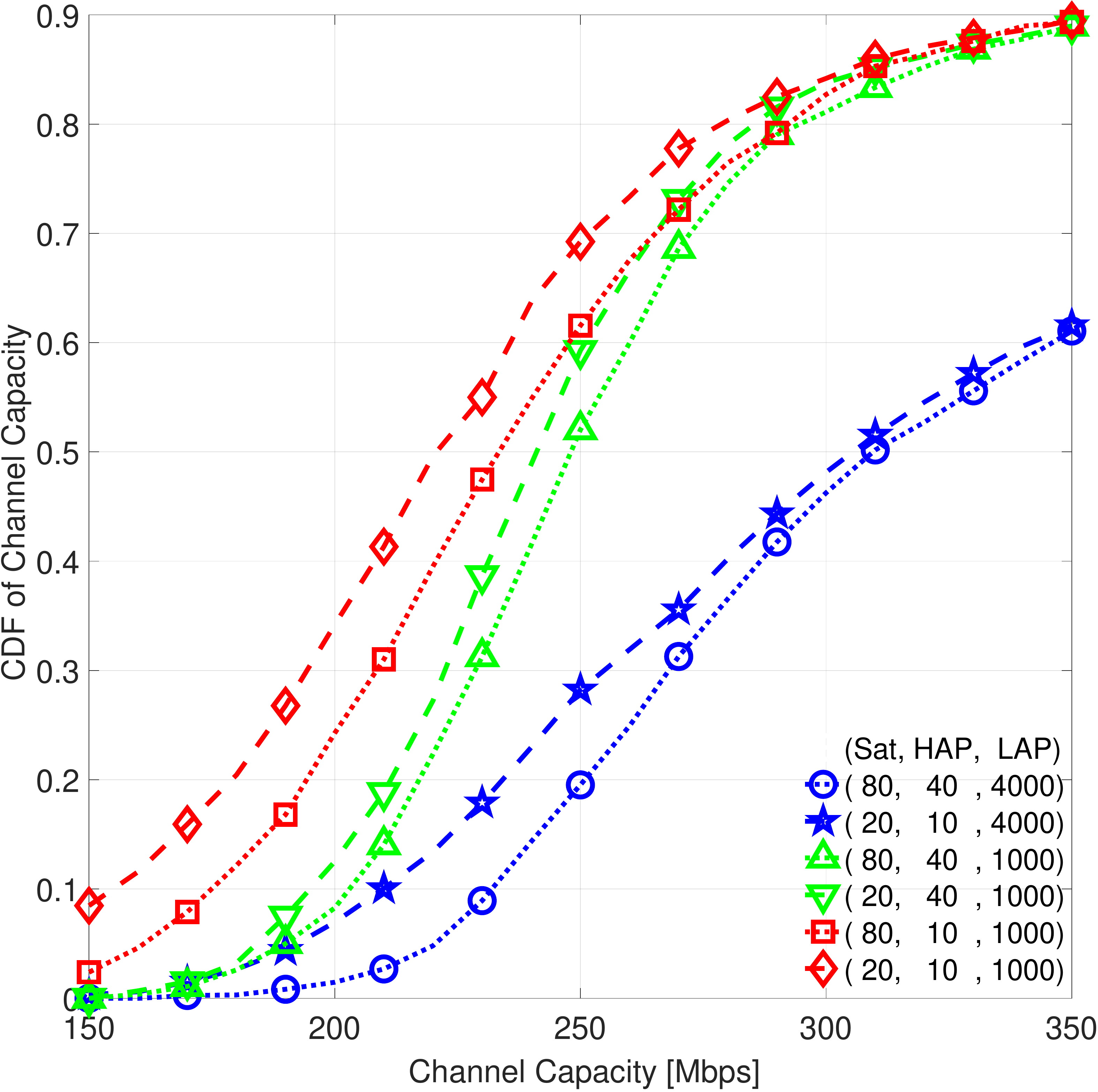}
    \label{fig:Capacity}}
    \subfigure[Energy Efficiency]{
    \includegraphics[width=0.43\linewidth]{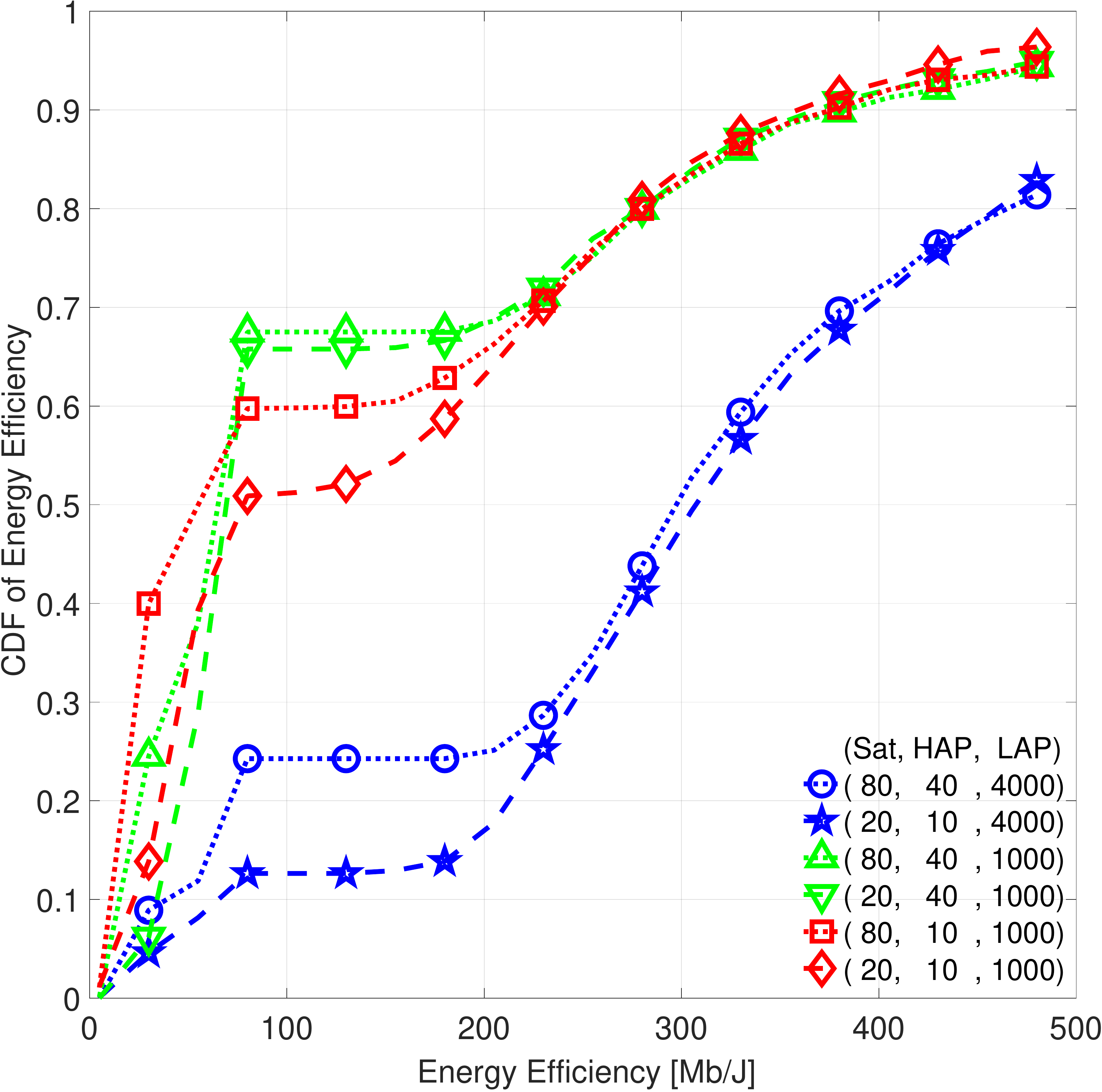}
    \label{fig:energy}}
    \caption{\revise{Capacity and energy efficiency analysis in
post-disaster reconstructed network.}}
\end{figure}

\begin{figure}
    \centering
    \includegraphics[width=0.5\linewidth]{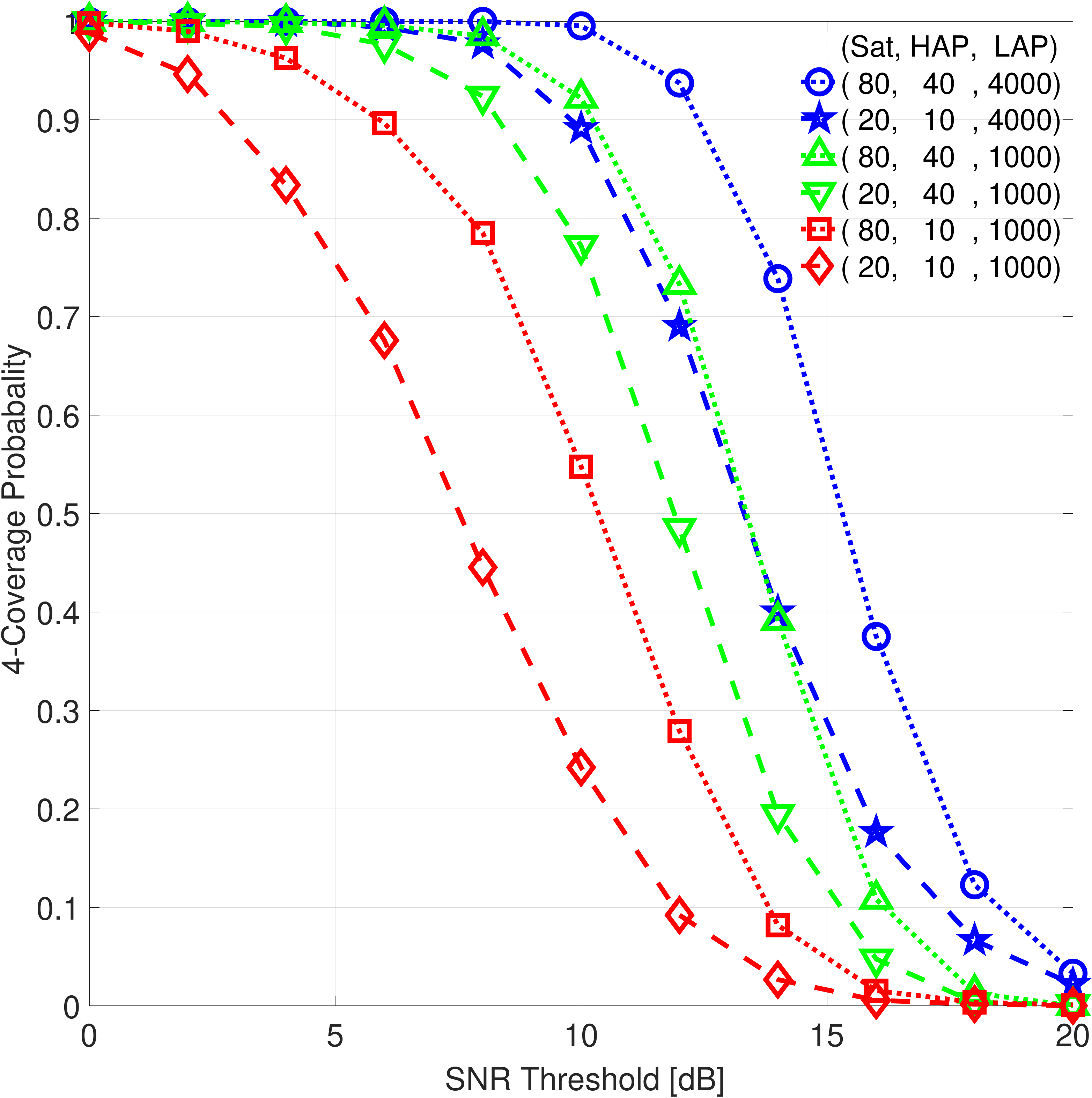}
    \caption{$k$-coverage probability under different thresholds ($k=4$).}
    \label{fig:kcoverage}
\end{figure}

Considering that availability is a key system-level metric in both relay communications and remote area networks, we \revise{constructed} a hybrid HAP and LEO satellite network to carry out an availability analysis. Ground users are modeled as a homogeneous PPP on a spherical cap with density $1/$km$^{2}$. The area of the spherical cap is $1.55 \times 10^5$~km$^2$, its dome angle is $\frac{1}{45}\pi$, and its radius is $6371$~km, which is equal to the Earth radius. 
\par

Assuming that the nearest terrestrial base station from a core network is $500$~km away from the center of the spherical cap so that there is no terrestrial base station serving this area. Therefore, users in a remote area cannot establish direct communication links with the core network and must be relayed by HAPs or LEO satellites. The satellites form a homogeneous spherical BPP in the whole sphere with a radius of 6921 km, which is the same radius as the recently deployed Starlink constellation. Unlike satellites, HAPs can be deployed stationary relative to the ground, thus we deploy HAPs directly above the user. HAPs are modeled by a homogeneous BPP on a spherical cap with a radius of 6391 km (the HAPs altitude is 20~km above Earth's mean sea level) and the same dome angle as users. 
\par

A randomly selected user in a remote area initiates a communication request. The relay communication link is available when there is at least one NTN platform in the common LoS region between the user and the ground base station. Fig.~\ref{fig:Remote} shows the ratio of the number of available communication links to the number of communication requests under the different numbers of HAPs and satellites. Even though LEO satellites are distributed on the whole sphere, they still contribute more to availability than HAP. Satellites have always been the first choice to promote equal access to networks. A small-scale LEO satellite constellation of $40$ satellites assisted by $50$ HAPs is sufficient to provide $95\%$ coverage over a region with $1.55 \times 10^5$~km$^2$ area.

\subsection{Capacity and Energy Efficiency of Post-disaster Networks}

We consider an aftermath scenario in which a ground network in a 2-dimensional plane with a radius of 30 km was destroyed by a disaster. The locations of the ground residents in this disaster area follow a homogeneous PPP with a density of 100/km$\rm ^{2}$. LAPs form a homogeneous BPP at an altitude of 80 m in the disaster area. HAPs and satellites follow the same distribution as subsection \ref{simulation1}. Therefore, the deployment area of HAP will be larger than the disaster area.

In this NTN system composed of LAPs, HAPs, and satellites, all the platforms transmit in an RF frequency of 2 GHz. LAPs adopt the same channel model as in \cite{qin2020performance}, and the channel model adopted by HAPs and satellites is given in \cite{MU_2}. The transmission powers of LAPs, HAPs, and satellites are 1 dBW, 8 dBW, and 15 dBW, respectively, and their antenna gains are 10 dB, 20 dB, and 70 dB, respectively. The small-scale fading parameters of the LAP channel are the same as \cite{qin2020performance}, and the small-scale fading parameters of the HAP and satellite channels are given in \cite{MU_2}. The environmental noise is set to -174 dBm/Hz.


As shown in the lines with circle and star marks in Fig.~\ref{fig:Capacity}, when the number of LAPs increases, the channel capacity increases significantly. The cumulative distribution function (CDF) curves corresponding to the red and green lines  converge, indicating that almost only LAPs can achieve a channel capacity greater than 300 Mbps. With the same number of LAPs, increasing the number of HAPs brings more benefits to channel capacity than increasing the number of satellites. The curves in Fig.~\ref{fig:energy} show the same convergence characteristics as in Fig.~\ref{fig:Capacity}, which means that only LAPs can achieve an energy efficiency of more than 250~Mb/J. The slope of the curves in the interval of $80-200$ Mb/J is steep, further indicating that the energy efficiency that HAP and satellites can achieve is lower than 80~Mb/J. When the number of UAVs is 1000, HAPs are more likely to be associated with satellites.



\subsection{$k$-Coverage Probability in Military \revise{Operations}}

This subsection investigates the $k$-coverage probability when $k=4$ during a military operation, that is, the probability that a military target in a given ground region is covered by at least four NTN platforms simultaneously. The distribution model and channel model for the three types of platforms in NTN are similar to those in the previous two subsections. It is assumed that the target is distributed in the same spherical cap as the users in subsection~\ref{simulation1}.

\par

Fig.~4 shows the simulation results of 4-coverage probability as a function of the SNR threshold. NTN platforms that are closer to the target can achieve better communication performance. The average SNR that LAP can provide is 10--20~dB, while the SNR of the HAP signal is greater than 5dB. When the military targets can demodulate signals with SNR over 5~dB, deploying more than 40 HAPs and 1000 LAPs over an area of $1.5-2 \times 10^5$km$^2$ can achieve 4-coverage for almost all military targets.


\section{Conclusion}
In this paper, we started with an introductory discussion about the concepts of NTNs and system-level metrics, and the motivation to apply SG tools. Then, we discussed suitable SG models for different platforms and compared their performance. Three typical communication issues, namely coverage, relay, and routing, were introduced and the system-level metrics of each issue were elaborated. The requirements for system-level metrics were further discussed for the three representative application scenarios of NTNs. Finally, we conducted some new combinations of system-level metrics with communication issues. The detailed SG-based analysis framework and numerical results of the metrics are presented.

\bibliographystyle{IEEEtran}
\bibliography{main}

\begin{thebibliography}{10}
\providecommand{\url}[1]{#1}
\csname url@samestyle\endcsname
\providecommand{\newblock}{\relax}
\providecommand{\bibinfo}[2]{#2}
\providecommand{\BIBentrySTDinterwordspacing}{\spaceskip=0pt\relax}
\providecommand{\BIBentryALTinterwordstretchfactor}{4}
\providecommand{\BIBentryALTinterwordspacing}{\spaceskip=\fontdimen2\font plus
\BIBentryALTinterwordstretchfactor\fontdimen3\font minus
  \fontdimen4\font\relax}
\providecommand{\BIBforeignlanguage}[2]{{%
\expandafter\ifx\csname l@#1\endcsname\relax
\typeout{** WARNING: IEEEtran.bst: No hyphenation pattern has been}%
\typeout{** loaded for the language `#1'. Using the pattern for}%
\typeout{** the default language instead.}%
\else
\language=\csname l@#1\endcsname
\fi
#2}}
\providecommand{\BIBdecl}{\relax}
\BIBdecl

\bibitem{belmekki2022unleashing}
B.~E.~Y. Belmekki \emph{et~al.}, ``Unleashing the potential of networked
  tethered flying platforms: {P}rospects, challenges, and applications,''
  \emph{IEEE Open Journal of Vehicular Technology}, vol.~3, pp. 278--320, 2022.

\bibitem{MU_2}
A.~Talgat \emph{et~al.}, ``Stochastic geometry-based analysis of {LEO}
  satellite communication systems,'' \emph{IEEE Communications Letters},
  vol.~25, no.~8, pp. 2458--2462, 2021.

\bibitem{OK_1}
N.~Okati \emph{et~al.}, ``Downlink coverage and rate analysis of low {E}arth
  orbit satellite constellations using stochastic geometry,'' \emph{IEEE
  Transactions on Communications}, vol.~68, no.~8, pp. 5120--5134, 2020.

\bibitem{lou2021green}
Z.~Lou \emph{et~al.}, ``Green tethered {UAV}s for {EMF}-aware cellular
  networks,'' \emph{IEEE Transactions on Green Communications and Networking},
  vol.~5, no.~4, pp. 1697--1711, 2021.

\bibitem{lee2022coverage}
J.~Lee \emph{et~al.}, ``Coverage analysis of {LEO} satellite downlink networks:
  Orbit geometry dependent approach,'' \emph{available online:
  https://arxiv.org/abs/2206.09382}, 2022.

\bibitem{choi2023cox}
C.-S. Choi \emph{et~al.}, ``Cox point processes for multi-altitude leo
  satellite networks,'' \emph{arXiv preprint arXiv:2301.02469}, 2023.

\bibitem{wang2022ultra}
R.~Wang \emph{et~al.}, ``Ultra-dense {LEO} satellite-based communication
  systems: {A} novel modeling technique,'' \emph{IEEE Communications Magazine},
  vol.~60, no.~4, pp. 25--31, 2022.

\bibitem{jung2022performance}
D.-H. Jung \emph{et~al.}, ``Performance analysis of satellite communication
  system under the shadowed-rician fading: A stochastic geometry approach,''
  \emph{IEEE Transactions on Communications}, vol.~70, no.~4, pp. 2707--2721,
  2022.

\bibitem{qin2020performance}
Y.~Qin \emph{et~al.}, ``Performance evaluation of {UAV}-enabled cellular
  networks with battery-limited drones,'' \emph{IEEE Communications Letters},
  vol.~24, no.~12, pp. 2664--2668, 2020.

\bibitem{al2021session}
A.~Al-Hourani, ``Session duration between handovers in dense {LEO} satellite
  networks,'' \emph{IEEE Wireless Communications Letters}, vol.~10, no.~12, pp.
  2810--2814, 2021.

\bibitem{belbase2018coverage}
K.~Belbase \emph{et~al.}, ``Coverage analysis of millimeter wave
  decode-and-forward networks with best relay selection,'' \emph{IEEE Access},
  vol.~6, pp. 22\,670--22\,683, 2018.

\bibitem{wang2022stochastic}
R.~Wang \emph{et~al.}, ``Stochastic geometry-based low latency routing in
  massive {LEO} satellite networks,'' \emph{IEEE Transactions on Aerospace and
  Electronic Systems}, pp. 1--14, 2022.

\bibitem{haenggi2005routing}
M.~Haenggi, ``On routing in random {R}ayleigh fading networks,'' \emph{IEEE
  Transactions on Wireless Communications}, vol.~4, no.~4, pp. 1553--1562,
  2005.

\bibitem{keeler2013sinr}
H.~P. Keeler \emph{et~al.}, ``{SINR-based} k-coverage probability in cellular
  networks with arbitrary shadowing,'' in \emph{International Symposium on
  Information Theory}.\hskip 1em plus 0.5em minus 0.4em\relax IEEE, 2013, pp.
  1167--1171.

\end{thebibliography}

\section*{biographies}
\begin{IEEEbiographynophoto}
{Qi Huang} received his B.Eng degree from SWPU, China in 2017, and is pursuing his master's degree in KAUST.
\end{IEEEbiographynophoto}
\begin{IEEEbiographynophoto}
{Baha Eddine Youcef Belmekki} received the M.Sc. degree in wireless communications and networking from the USTHB, Algeria, in 2013, and the Ph.D. degrees in wireless communications from INPT, France, in 2020. He is currently a postdoctoral fellow at KAUST.
\end{IEEEbiographynophoto}
\begin{IEEEbiographynophoto}
{Ahmed M. Eltawil} received his Ph.D. degree from UCLA in 2003. He was with UCI since 2015 as a professor and now is a professor of electrical engineering at KAUST.
\end{IEEEbiographynophoto}
\begin{IEEEbiographynophoto}
{Mohamed-Slim Alouini} received his Ph.D. degree in electrical engineering from the California Institute of Technology, Pasadena, in 1998. He joined KAUST as a professor of electrical engineering in 2009.
\end{IEEEbiographynophoto}

\end{document}